\documentclass{article}
\usepackage{amsmath,graphicx}
\usepackage[preprint]{spconf}
\usepackage[utf8]{inputenc}
\usepackage[T1]{fontenc}
\usepackage{fancyhdr}

\copyrightnotice{\copyright this is copyright}
\title{BIOLOGICALLY INSPIRED SPEECH EMOTION RECOGNITION}

\name{Reza Lotfidereshgi, Philippe Gournay}
\address{Speech and Audio Research Group\\
Université de Sherbrooke\\
Sherbrooke (Québec) J1K 2R1 Canada}
\begin{document}
%

\maketitle
\copyrightnotice{\copyright 2017 IEEE. Personal use of this material is permitted. Permission from IEEE must be obtained for all other uses. }

\begin{abstract}
Conventional feature-based classification methods do not apply well to automatic recognition of speech emotions, mostly because the precise set of spectral and prosodic features that is required to identify the emotional state of a speaker has not been determined yet. This paper presents a method that operates directly on the speech signal, thus avoiding the problematic step of feature extraction. Furthermore, this method combines the strengths of the classical source-filter model of human speech production with those of the recently introduced liquid state machine (LSM), a biologically-inspired spiking neural network (SNN). The source and vocal tract components of the speech signal are first separated and converted into perceptually relevant spectral representations. These representations are then processed separately by two reservoirs of neurons. The output of each reservoir is reduced in dimensionality and fed to a final classifier. This method is shown to provide very good classification performance on the Berlin Database of Emotional Speech (Emo-DB). This seems a very promising framework for solving efficiently many other problems in speech processing.
\end{abstract}
\begin{keywords}
speech emotion recognition, source-filter model, liquid state machine, reservoir computing
\end{keywords}
\section{Introduction}
\label{sec:intro}


Speech is a fundamental means of communicating not only words, but also a vast range of human emotions. Consequently, speech processing applications, such as human-machine interfacing and speech recognition, could benefit from the introduction of a reliable method for automatic recognition of human emotions through speech.

Conventional speech emotion recognition methods consist of a feature extraction step followed by a classifier. Various spectral and prosodic features can be used \cite{rao2013robust}. Finding the “best” set of features,  namely, one that is both complete and compact, is a critical step which has a considerable impact on the performance of the system. State of the art conventional methods mostly differ in their choice of features and of classifier type \cite{el2011survey,anagnostopoulos2015features}.

In recent years, there has been an increasing trend toward developing speech processing methods that operate directly on the speech signal in order to avoid the problematic feature extraction step. For example, Convolutional Neural Networks (CNNs) \cite{trigeorgis2016adieu} and Deep Neural Networks (DNNs) \cite{fayek2015towards}, have been successfully used for recognizing emotions directly from raw temporal or spectral data.

The Liquid State Machine (LSM) is another recently proposed method that operates directly on raw data. The LSM relies on a network of spiking neurons that are much closer to biological neurons than the rate-based model used in CNNs and DNNs. Despite its theoretical appeal, the LSM is slow in finding practical applications. The main problem when implementing an LSM is to create a specific reservoir design that is best adapted to the task at hand  \cite{lukovsevivcius2009reservoir}. In this paper, this problem is solved by introducing prior knowledge about the human speech production system into the LSM. Without any loss in terms of information, the speech signal is divided into two components: the source and the vocal tract. Individually, each component is easier to process by a reservoir of spiking neurons. Furthermore, the inclusion of a production model in the recognition system is justified by the motor theory of speech perception, that states that people perceive speech by identifying the vocal tract gestures that produced it \cite{liberman1967perception}.

The outline of the paper is as follows. The source-filter model for human speech production and the principles underlying the liquid state machine are reviewed in section 2. The proposed biologically inspired method is presented in section 3. Some experimental results are given and discussed in section 4, and conclusions are drawn in section 5.

\section{Relation to prior work}
\label{sec:method}

\subsection{The source-filter speech production model}
\label{ssec:Source-filter}

According to the source-filter model of human speech production, a speech signal is produced by passing a source of air pressure through an acoustic filter \cite{fant1971acoustic}. The source is a combination of a noise-like turbulent excitation produced by constrictions along the vocal tract (for unvoiced speech) and a quasi-periodic excitation produced by vibrating vocal folds (for voiced speech). The filter represents the variable response of the vocal tract.
In practice, the most commonly used method to separate the contributions of the source and the filter is the Linear Predictive (LP) analysis. The LP analysis is a frame-based process which results in: (1) a set of LP coefficients which represent the filter for the frame; and (2) a residual error signal which represents the source. Equation \ref{eq:1} shows the calculation of a predicted speech sample $\widetilde{x}(n)$ from past speech samples $x(n-i)$ and the calculation of a residual sample $e(n)$. The Levinson-Durbin algorithm is usually used to find the $a_{i}$ coefficients that minimize the quadratic error $E$ as shown in equation \ref{eq:1_1}.
\begin{equation}\label{eq:1}
\widetilde{x}(n)=\sum_{i=1}^{M}a_{i}x(n-i),\,\,\,\,\,\,\,  e(n)=x(n)-\widetilde{x}(n)
\end{equation}

\begin{equation}\label{eq:1_1}
E=\sum_{n}^{ }e(n)^2
\end{equation}

The LPC analysis has long proven to be a very efficient tool in speech processing and is now used for example in every speech coder.

\subsection{The Liquid State Machine}
\label{ssec:reservoir}

A reservoir computing system consists of a Recurrent Neural Network (RNN) followed by an output layer of neurons that performs the final recognition/classification task \cite{lukovsevivcius2009reservoir}. In a reservoir computing system, the RNN is randomly created and does not need to be trained using supervised methods such as the gradient descent. The output layer, in contrast, is trained using a supervised method. Reservoir computing is successful for complex nonlinear classification tasks for two reasons. First, because training an RNN using a gradient-descent algorithm would be time consuming and prone to convergence issues. Secondly, because reservoir computing has been shown to outperform most other nonlinear identification, prediction and classification methods on various problems.

\begin{figure}

  \centerline{\includegraphics[width=8.0cm]{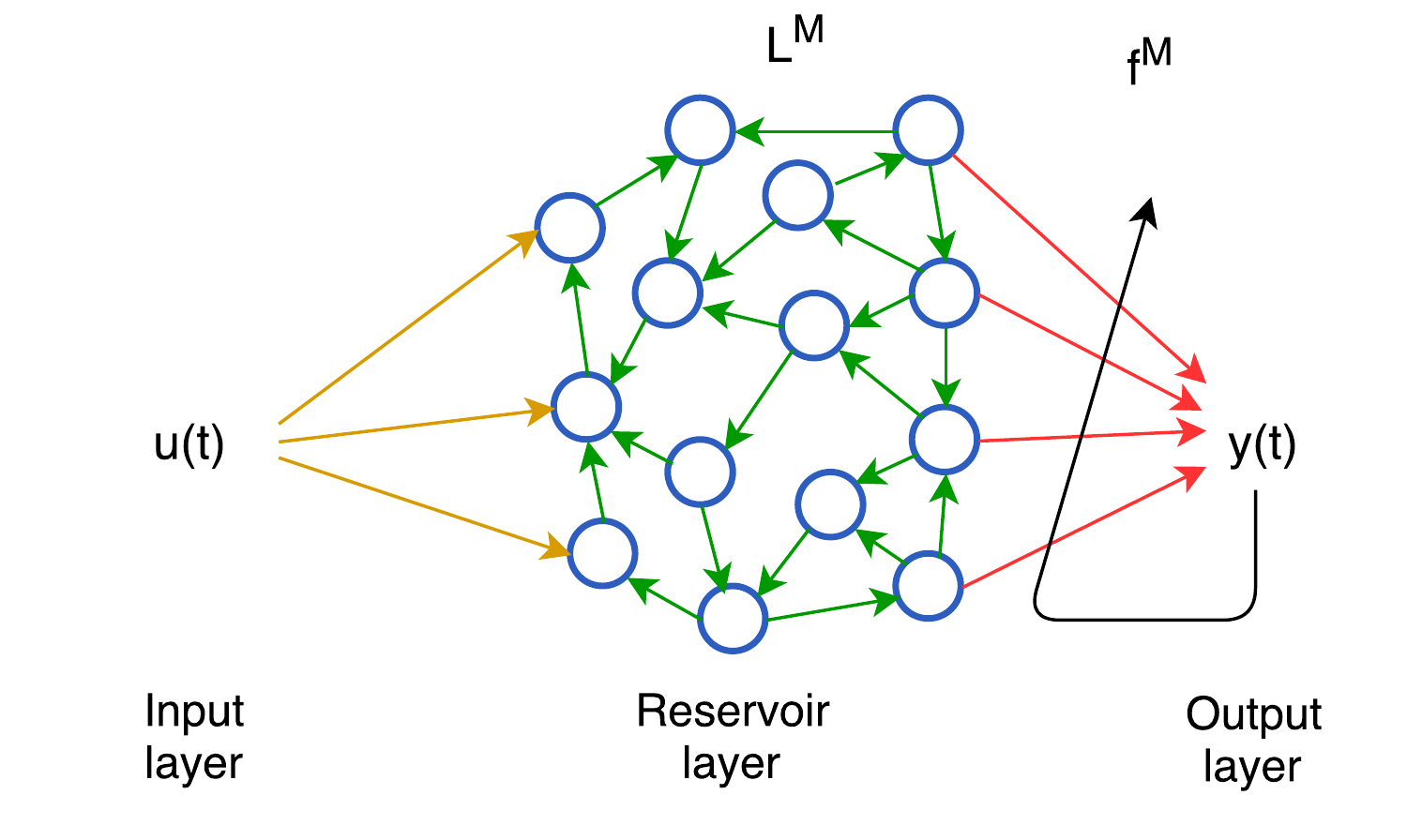}}
%
\caption{A typical LSM structure. Only the output layer is trained using a supervised methods.}
\label{fig:res}
\end{figure}

The Liquid State Machine (LSM) is a special type of reservoir computing method where the reservoir is a Spiking Neural Network (SNN) \cite{maass2002model}. SNNs use temporal coding and therefore process information in very much the same way as a biological neural structure does. Fig.\ref{fig:res} shows a typical LSM structure. First, the LSM uses a function $L^{M}$ to map the input $u(t)$ to the “liquid state” $x(t)$, where $x(t)$ is an arbitrary nonlinear function of the input $u(t)$ and of the past input values. Secondly, a memoryless function $f^{M}$ maps $x(t)$ to the output $y(t)$. This “readout function” is trained for the task to accomplish. The SNN performs a nonlinear mapping from the input space to the high dimensional “liquid state” space. As a result of this projection, the separation of different classes by the readout function is much easier. Several methods including Support Vector Machine (SVM), Multi Layer Perceptron (MLP) and ridge regression have been tried as reservoir readouts \cite{lukovsevivcius2009reservoir}.


\section{Proposed method}
\label{ssec:proposed}

\begin{figure}[t]
    \centering
\includegraphics[width=7cm]{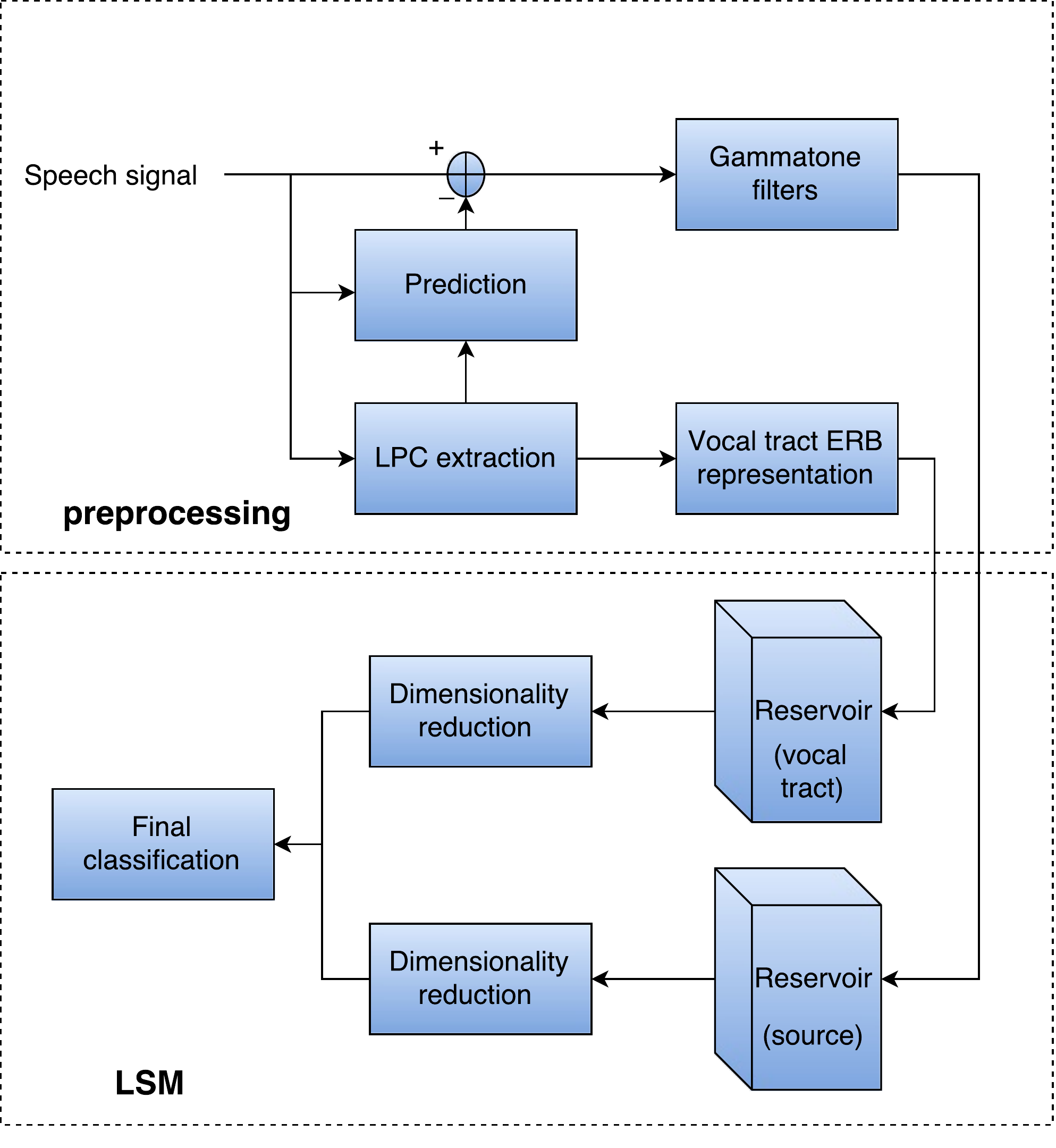}
\caption{Flowchart of proposed emotion recognition method.}
\label{fig:diag}
\end{figure}

Fig.\ref{fig:diag} shows the flowchart of the proposed speech emotion recognition method. The recognition process is divided into two steps: the preprocessing and the liquid state machine.

In the preprocessing step, the input speech signal is divided into two orthogonal and complementary components that are transformed and perceptually shaped according to the properties of the human cochlea. Specifically, an LP analysis is performed on a frame base. The prediction residual is calculated according to equation 1 and decomposed using a 77-channel gammatone filterbank with ERB scaling. This constitutes the input of the first reservoir. In parallel, the frequency response of each all-pole LP filter is computed to reveal the formant structure of the speech signal. This frequency response is also shaped using the exact same ERB scaling and constitutes the input of the second reservoir.

As in the lower auditory nuclei, even auditory cortex has the tonotopic structure \cite{munkong2008auditory}. Such structure suggests that closer frequency channels are processed by closer groups of neurons. In the design of the reservoirs, the neurons are therefore arranged in 3D structures, each reservoir containing 77 layers of 3*3 neurons. Each layer of neurons is excited by only one of the 77 input channels, in order of increasing frequency. Connections between closer neurons are favored, with a probability of connection between neuron $n1$ and $n2$ that depends on the distance $D(n1, n2)$ according to equation 3. Parameters $C$ and $\lambda$ are responsible for controlling the reach and density of the connections, and are set respectively to 1 and 3.4. These values were determined after some experiments and could probably be further optimized.

\begin{equation}\label{eq:2}
P(n1,n2)=Ce^{-\frac{D^{2}(n1,n2))}{\lambda ^{2}}}
\end{equation}

A standard implementation of the integrate-and-fire neuron by Troyer is used \cite{troyer1997integrate}. The Asymmetric Spike Time-Dependent Plasticity (STDP) is then used as the learning rule to adapt the conductance of the synapses throughout the speech sample. This learning rule is known to result in stable networks that are very effective at extracting the correlations present in the input \cite{song2001cortical}. The exact learning rule is given in equation \ref{eq:3}.

\begin{equation}\label{eq:3}
f\left ( \Delta  \right )=\left\{\begin{matrix} A_{+}e^{\frac{\Delta }{\tau _{+}}} \begin{matrix}, & \Delta<0\end{matrix}  \\ -A_{-}e^-{\frac{\Delta }{\tau _{-}}},  \Delta>0 \end{matrix}\right.
\end{equation}

More details about this learning rule can be found in \cite{song2001cortical}. $ \Delta$ is the time difference between pre- and post-synaptic spikes. $A_{+}$ and $A_{-}$ are maximum amount of synaptic modifications. Two key parameters to be set are the time constants $\tau_{+}$ and $\tau_{-}$ because they condition the memory of the reservoir. Following \cite{song2001cortical}, $\tau_{+}$ was set to 20\,ms and $\tau_{-}$ was tuned to maximize performance (see section 4). The simulation of neural activity is done using the Brian2 simulator \cite{stimberg2014equation}.

To reduce dimensionality, Principal Component Analysis (PCA) is applied to the average activity of the neurons from each reservoir. Compared to the widely used ridge regression, PCA presents the advantage of being able to shrink the output of the two reservoirs separately. The outputs of the two PCAs are simply combined. For final recognition, Linear Discriminant Analysis (LDA) is used.

\begin{figure}[t]
\begin{minipage}[b]{1.0\linewidth}
  \centering
  \centerline{\includegraphics[width=8.5cm]{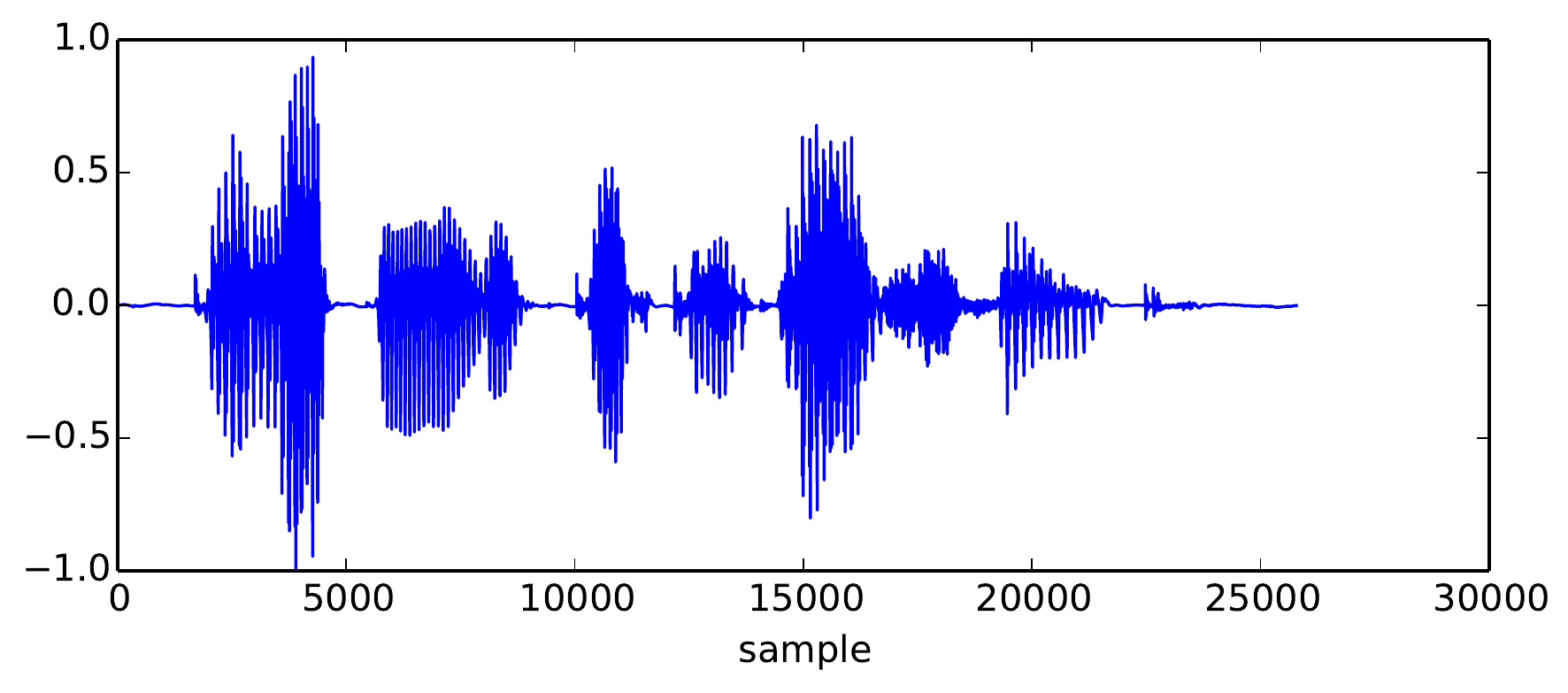}}
  \centerline{(a) sample speech signal}\medskip
\end{minipage}

\begin{minipage}[b]{0.48\linewidth}
  \centering
  \centerline{\includegraphics[width=4cm]{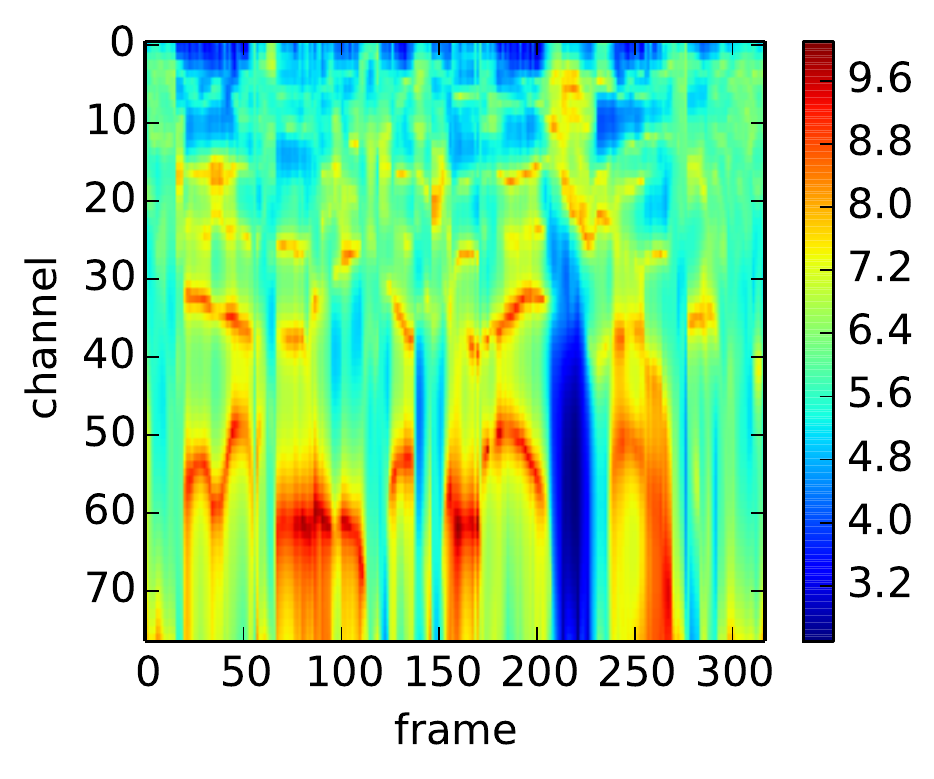}}
  \centerline{(b)  vocal tract contribution }\medskip
\end{minipage}
\hfill
\begin{minipage}[b]{0.48\linewidth}
  \centering
  \centerline{\includegraphics[width=4cm]{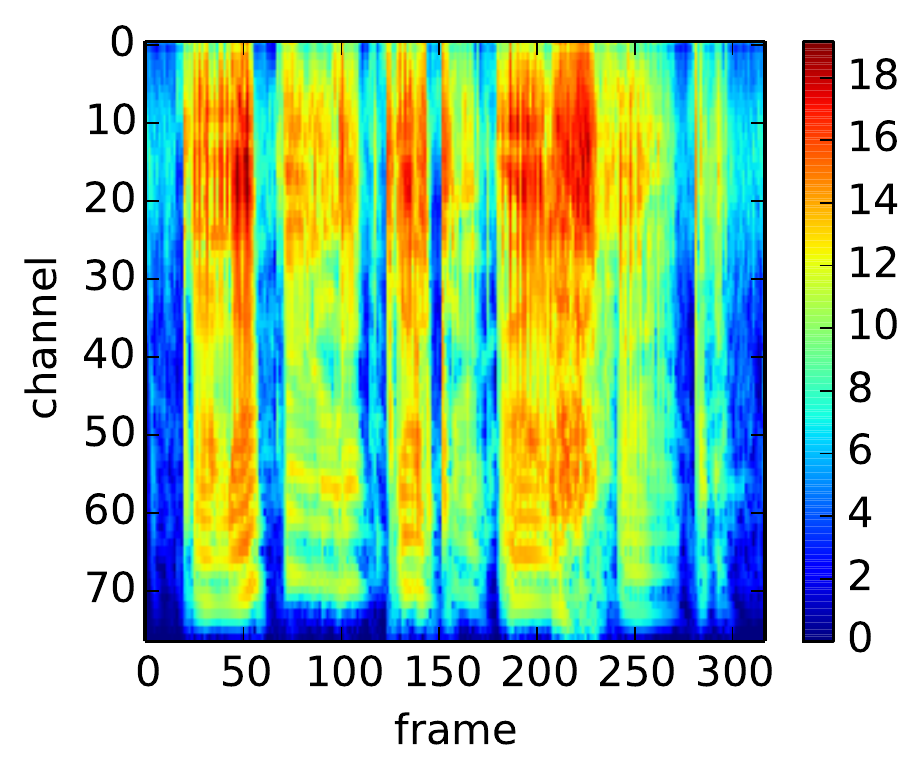}}
  \centerline{(c)  source contribution }\medskip
\end{minipage}
\caption{Preprocessing of speech signal. The scales of the spectro-temporal representations are in dB.}
\label{fig:speech}
\end{figure}

\section{Experiments}
\label{sec:experinents}

\subsection{Berlin database of emotional speech}
\label{ssec:berlin}

The proposed method was tested on the Berlin database of emotional speech (Emo-DB, \cite{burkhardt2005database}). This is a well recorded and now widely used emotional speech database. It is easily accessible and well documented. It contains 535 utterances produced by ten professional actors pronouncing ten different texts and covers seven different emotions.

\subsection{preprocessing}
\label{ssec:preprocessing-ex}

The preprocessing step first consists in an LP analysis of the input speech signal. The autocorrelation method is used to estimate LP filters of order 16. A 30\,ms Hamming window is used so that the formant structure is adequately captured. The LP coefficients are updated every 5\,ms in order to closely track the changes in the vocal tract.
The source and vocal tract components of the speech signal are then separated. First, the LP residual is computed and fed to a 77-channel gammatone filterbank. For each channel of the filterbank, the energy of 5\,ms segments is computed and a logarithm is applied to reduce the dynamic range of this representation of the source component. Secondly, the frequency response of each LP filters is computed and shaped using an ERB frequency scaling. A logarithm is also applied to reduce the dynamic range of this representation.

An example of emotional speech signal is represented in Fig.\ref{fig:speech}(a). The corresponding source and vocal tract representations are presented in  Fig.\ref{fig:speech}(b) and Fig.\ref{fig:speech}(c), respectively. These two spectro-temporal representations of the speech signal are used as inputs for two reservoirs of spiking neurons.

\begin{figure}[t]
  \centering
\includegraphics[width=7.3cm]{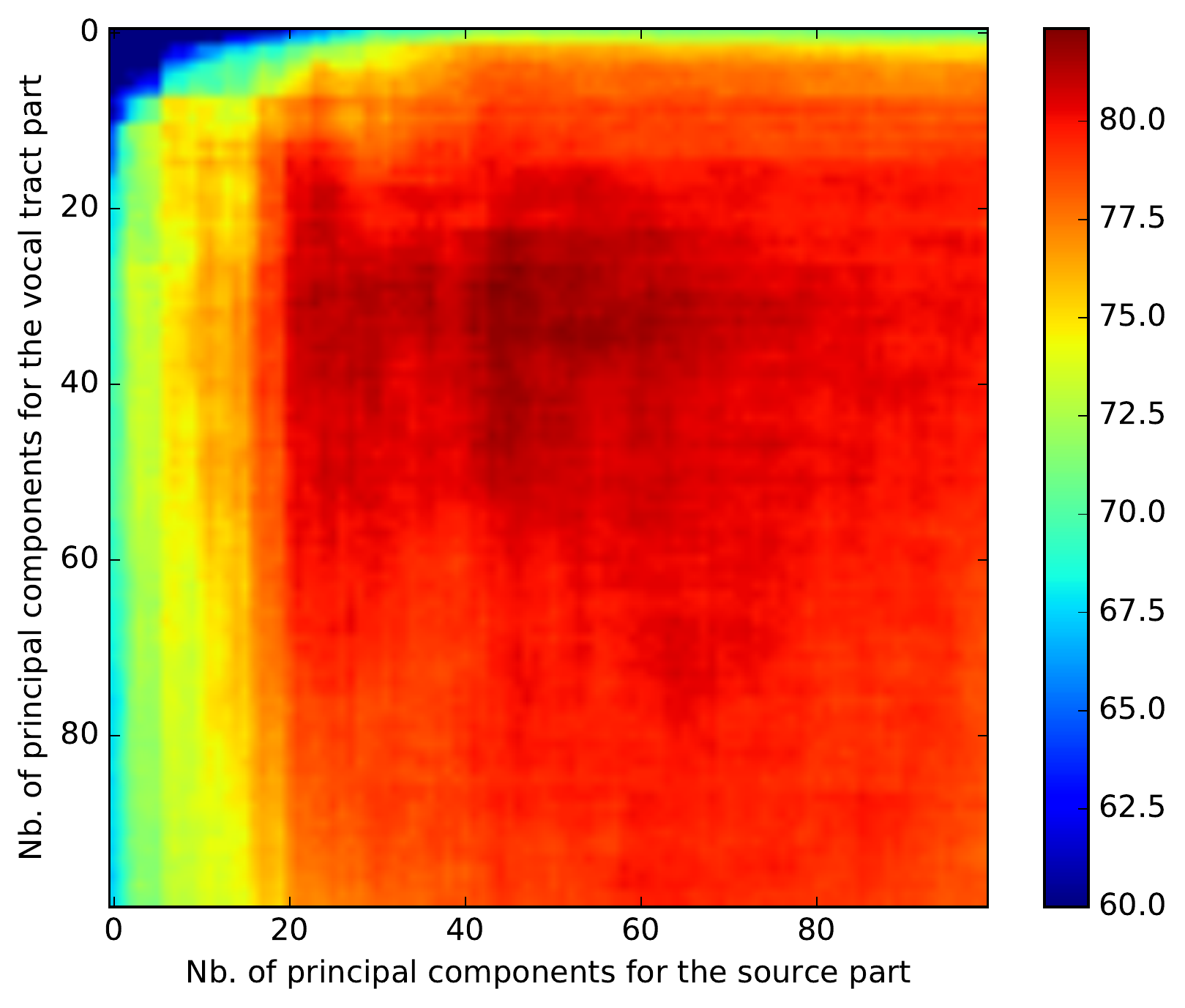}
\caption{Performance (in percent) of the proposed method for different numbers of principal components for each reservoir.}
\label{fig:performance}
\end{figure}

\subsection{LSM tuning}
\label{ssec:synamics}

 Fig.\ref{fig:performance} shows the recognition rate of the proposed method for different numbers of principal components for each of the two reservoirs. The reservoir for the vocal tract component was tuned for $\frac{\tau_{-}}{\tau_{+}} = 5$ and the reservoir for the source component was tuned for $\frac{\tau_{-}}{\tau_{+}} = 3$. The results are obtained using 50-fold cross validation where 90\% of the database is used for training and 10\% for testing. Results below 60\% of recognition rate are not shown. The vertical and horizontal axes are the number of principal component selected from the vocal tract and source reservoirs, respectively. The borders of the figure shows the performance when only one reservoir is used (no component from the other reservoir is selected). It is quite clear that both reservoirs contribute highly to the final recognition rate. The performance of the proposed method is not very sensitive to the choice of numbers of principal components, since the recognition rate stays above 80\% for a wide range of numbers of components.

The highest recognition rate of 82.35\% is achieved for 29 and 44 principal components for the vocal tract and the source reservoirs, respectively, and the 95\% confidence interval is $\pm 1.36\%$. Table 1 shows the corresponding confusion matrix.

Table \ref{Comp} compares the recognition rate obtained with the proposed method to those obtained with other methods that have been tested on the same emotional speech database. Using feature selection and fusion, the method presented by Jin in \cite{jin2014feature} achieved 83.10\% of correct recognition. It should be noted however that this method was tested on a subset of only 494 speech samples out of 535, which artificially increases the performance. Using an enhanced kernel isomap, the method presented by Zhang in \cite{zhang2013speech} achieved a recognition rate of 80.85\%. Finally, using rhyme and temporal features, the one presented by Bhargava in \cite{bhargava2013improving} achieved 80.60\%. With a recognition rate of 82.35\%, the method proposed in this paper compares favorably to these state of the art methods.

In another experiment, we used an LSM with one single reservoir to recognize emotion directly from the speech signal, without separating the source from the vocal tract. The same preprocessing as for the source component was used. The design of the rest of the system was not changed. After tuning the reservoir, a recognition rate of 75.73\% was obtained. The 6.62\% difference in recognition rate clearly indicates that including a source-filter model in the recognition system significantly improves performance.


\begin{table}[t]
\centering
\caption{Confusion matrix.}
\label{conf}
\includegraphics[width=6.3cm]{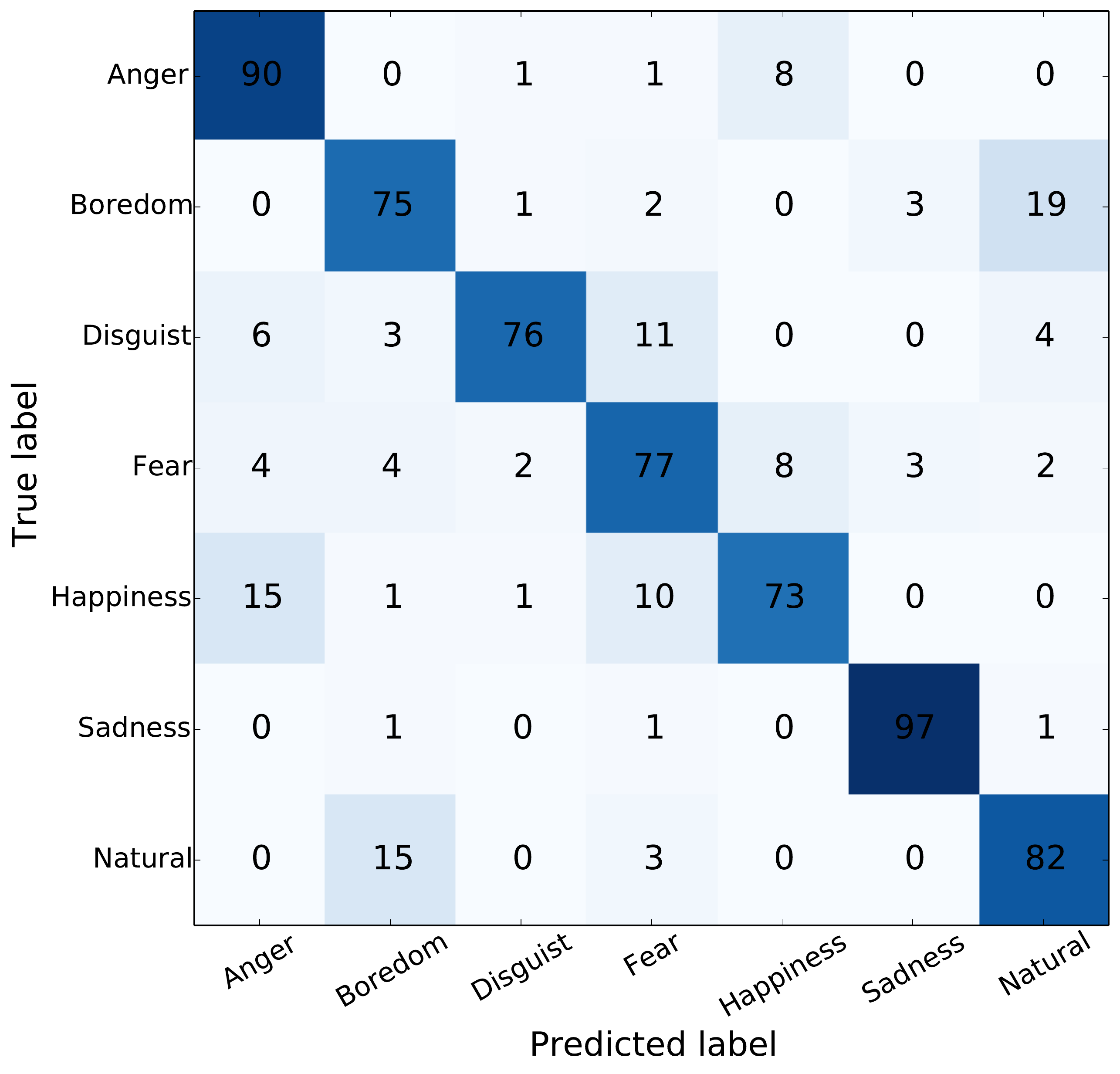}
\end{table}

\begin{table}[t]
\centering
\caption{Recognition rate compared with other methods.}
\label{Comp}
\begin{tabular}{|c|c|c|c|}
\hline
Our method & Jin & Zhang & Bhargava \\ \hline
82.35\% & *83.10\% & 80.85\% & 80.60\% \\ \hline
\multicolumn{3}{l}{* For a subset of Berlin Database}
\end{tabular}
\end{table}

\section{conclusions}
\label{sec:conclusions}
This paper proposed a new method for automatic recognition of speech emotions based on the Liquide State Machine (LSM), an emerging and very promising tool. This method operates directly on the speech signal and thus requires no feature extraction. It is based on several biological elements. First, the LSM includes a reservoir of spiking neurons which are very close to biological cortical neurons. Then, its original LSM design with two separate reservoirs (one for the source signal and the other for the vocal tract) builds upon the motor theory of human speech perception. This design is more flexible and tunable, as for example the size and memory of the two reservoirs can be tuned separately. One could imagine decomposing the signal even further, using for example rapidly evolving and slowly evolving waveform decomposition of the source signal \cite{choy1998waveform} Finally, the source and vocal tract components of the speech signal are both analyzed on an Equivalent Rectangular Bandwidth (ERB) scale which is a good model for the human peripheral auditory system.

The experimental results showed that this method provides a very good classification performance for an emotion recognition task. It is, however, a very general framework that should also perform well for many other speech processing tasks.

\vfill\pagebreak
\bibliographystyle{IEEEbib}
\bibliography{lukovsevivcius2009reservoir,rao2013robust,jin2014feature,maass2002model,burkhardt2005database,song2001cortical,el2011survey,zhang2013speech,bhargava2013improving,hastie2009elements,trigeorgis2016adieu,fayek2015towards,anagnostopoulos2015features,troyer1997integrate,stimberg2014equation,liberman1967perception,fant1971acoustic,munkong2008auditory,choy1998waveform}
\end{document}